\documentclass[final] {aipproc}
\pdfoutput=1
\layoutstyle{6x9}
\usepackage{graphicx}
\usepackage{slashed}
\usepackage{latexsym}
\usepackage{amsmath}

\begin{document}

\title{Leptogenic Supersymmetry at the LHC}


\author{Andrea De Simone}{
  address={ Center for Theoretical Physics,
        Massachusetts Institute of Technology, Cambridge, MA 02139}          
}

\begin{flushright}
\begin{footnotesize} MIT-CTP-4075 \end{footnotesize}
\end{flushright}

\begin{abstract}
Leptogenic Supersymmetry is a scenario characterized by copious lepton production in cascade decays. 
Due to the  high lepton multiplicity and the lack of significant missing energy, leptogenic supersymmetry provides very 
clean channels which can be probed already with the early LHC data.
Furthermore, the Higgs may be discovered in the $h\to b \bar b$ mode  because the leptons accompanying 
Higgs production efficiently suppress the background. 
\end{abstract}

\maketitle


\section{Introduction}

What kind of supersymmetric physics can be probed with  the first data collected at the LHC? Unfortunately, in most supersymmetric
scenarios it is difficult to disentangle the signals from the background and have statistically significant excesses of events with as little as few hundreds of pb$^{-1}$ of integrated luminosity. An exception is ``leptogenic supersymmetry''  which is characterized, as the name
suggests, by an abundant production of leptons in cascade decays.
Since events with many leptons are easily detected and have very small background, 
leptogenic supersymmetry is a particularly clean situation and within the reach of the very first data from LHC.

In this talk I will present the features of leptogenic supersymmetry and the main results of the analysis of two relevant channels.


\section{Main Features and Phenomenology}

Suppose the spectrum of superpartner masses is such that
\begin{equation}
m_{\tilde{g}}, m_{\tilde{q}}> m_{\tilde{\chi}^0},m_{\tilde{\chi}^{\pm}} > m_{\tilde{\ell}_L} > m_h, m_{\tilde{{\ell}}_R}\,,
\label{hierarchy}
\end{equation}
where the gluino can be either lighter or heavier than the squarks.
Spectra like this emerge in several scenarios of supersymmetry breaking, e.g. low-scale gaugino mediation \cite{lgm},  
gauge mediation \cite{gmsb} with a large number of messengers, models with Dirac gaugino masses \cite{Fox:2002bu}. 
I will not further discuss  how this hierarchy of masses can get generated at high energies, so no assumptions are made about its origin. 
I instead take a model-independent approach and study the phenomenological implications of the spectra described by (\ref{hierarchy}).

The decay kinematics ensures that multiple leptons are produced in every decay chain.
The production of new particles is dominated by QCD production of squarks and gluinos, which are assumed to be at the top of the mass spectrum. 
The production cross-section of squark pairs is around 1 pb, at 14 TeV center-of-mass energy and for TeV squarks.  
These heavy coloured particles decay into lighter charginos and neutralinos, emitting jets. 
The charginos and neutralinos are heavier than the sleptons and therefore decay into leptons and sleptons. All sleptons decay into the lightest slepton which is the  next-to-lightest supersymmetric particle (NLSP). The NLSP is assumed to be collider stable and eventually decays into the gravitino outside the detectors. The situation described above has been named ``leptogenic supersymmetry'' \cite{lepto}.

The typical final states are  characterized by two hard jets (produced by the decays of very energetic squarks),  two collider-stable sleptons, and at least two leptons. 
The number of leptons in the final state depends on whether the intermediate 
neutralinos or charginos go through a short ($\tilde\chi\to\tilde \ell_R \ell$) or a long ($\tilde\chi\to\tilde \ell_L\ell\to \tilde\ell_R\ell\ell'\ell''$)
decay chain.
Furthermore, heavy collider-stable particles with sufficient velocity ($\beta>0.95$) appear as muons with a delayed arrival at the muon chambers \cite{cmstdr, atlastdr}.
If the sleptons are misidentified as muons  then the final state shows four or more leptons and two hard jets.
These channels are new for SUSY searches: they are practically background-free, with almost no missing energy and 
they allow the mass reconstruction of most sparticles at low luminosity ($<1\, \textrm{fb}^{-1}$). 
In addition, the Standard Model-like Higgs boson can be discovered in the $h\to b\bar b$ mode.

\begin{figure}[t]
\includegraphics[width=5.7cm, height=3.5cm]{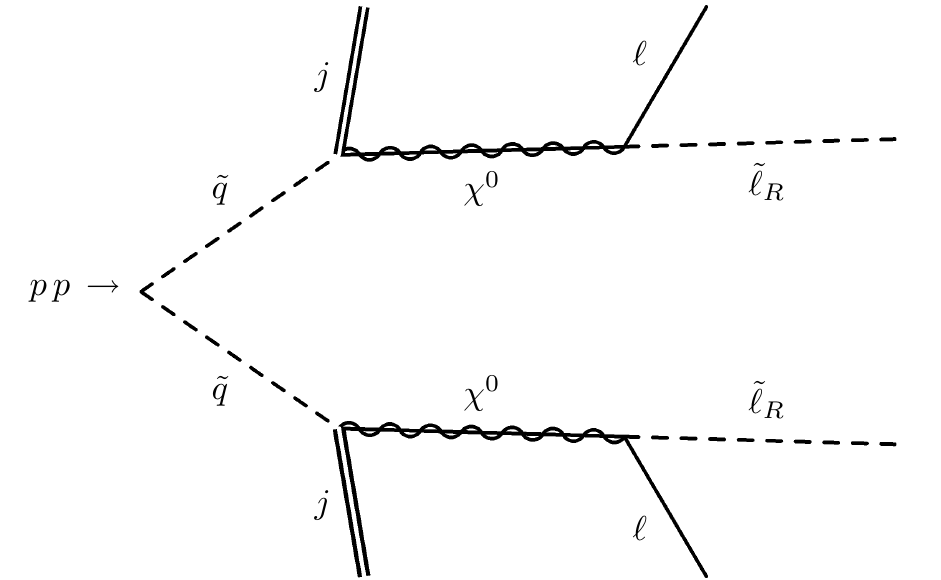}
\hspace{1cm}
\includegraphics[width=5.7cm, height=3.7cm]{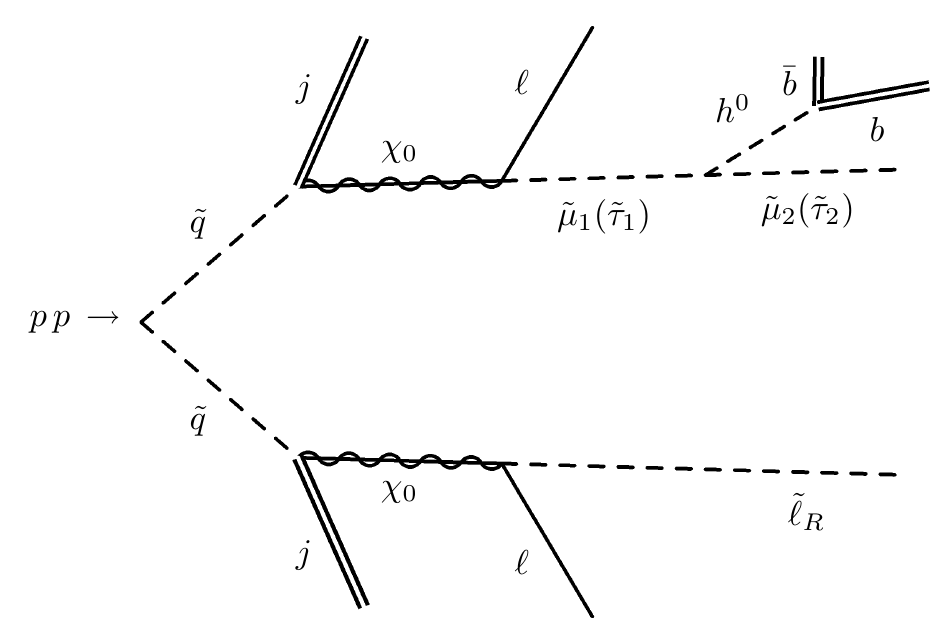}
\caption{\textit{Left:} Four-lepton channel. \textit{Right:} Channel for Higgs mass reconstruction.}
\label{decays} 
\end{figure}

\subsection{Four-lepton channel}

The four-lepton channel originates from the events  where neutralinos decay directly to stable sleptons (as depicted in Fig.~\ref{decays}).
We select four-lepton events asking for $n_{\ell}=4, n_{\rm{jet}}\geq 2$ with standard cuts $|\eta_{\rm{\ell}, \rm{jet}}|<2.5, p_T^{\rm{\ell}}> 10$ GeV, $p_T^{\rm{jet}}>15$ GeV, and the isolation cuts $\Delta R_{\rm{jj}, \ell\ell, \rm{j}\ell}>0.4$.
After applying these cuts, all SM backgrounds were estimated to be below the fb level. 
For the sample spectrum we considered, the production cross-section for this process is 220 (690) fb at 10 (14) TeV, after the cuts described above.

There is no missing energy in this channel, so the events can be fully reconstructed. In the absence of SM background, combinatorial background is the main obstacle to the reconstruction. 
Depending on whether sleptons are identified or not, we use different strategies to reconstruct the neutralinos  participating in the process.
Combinatorics can be efficiently reduced by pairing objects coming according to minimal $\Delta$R separation.
For the events with identified sleptons,  we pair a slepton with a nearby lepton through $\Delta$R$_{\ell \tilde{\ell}}$ selection.  We follow the same procedure to pair the slepton-lepton pair with the nearest jet. 
For the events with sleptons misidentified as muon, we form dilepton invariant masses and selected dilepton pairs with opposite charge and smaller $\Delta$R$_{\ell \ell}$. This $\Delta$R discrimination turned out to be  very powerful.

\begin{figure}[t]
  \includegraphics[width=4.8cm, height=5.5cm]{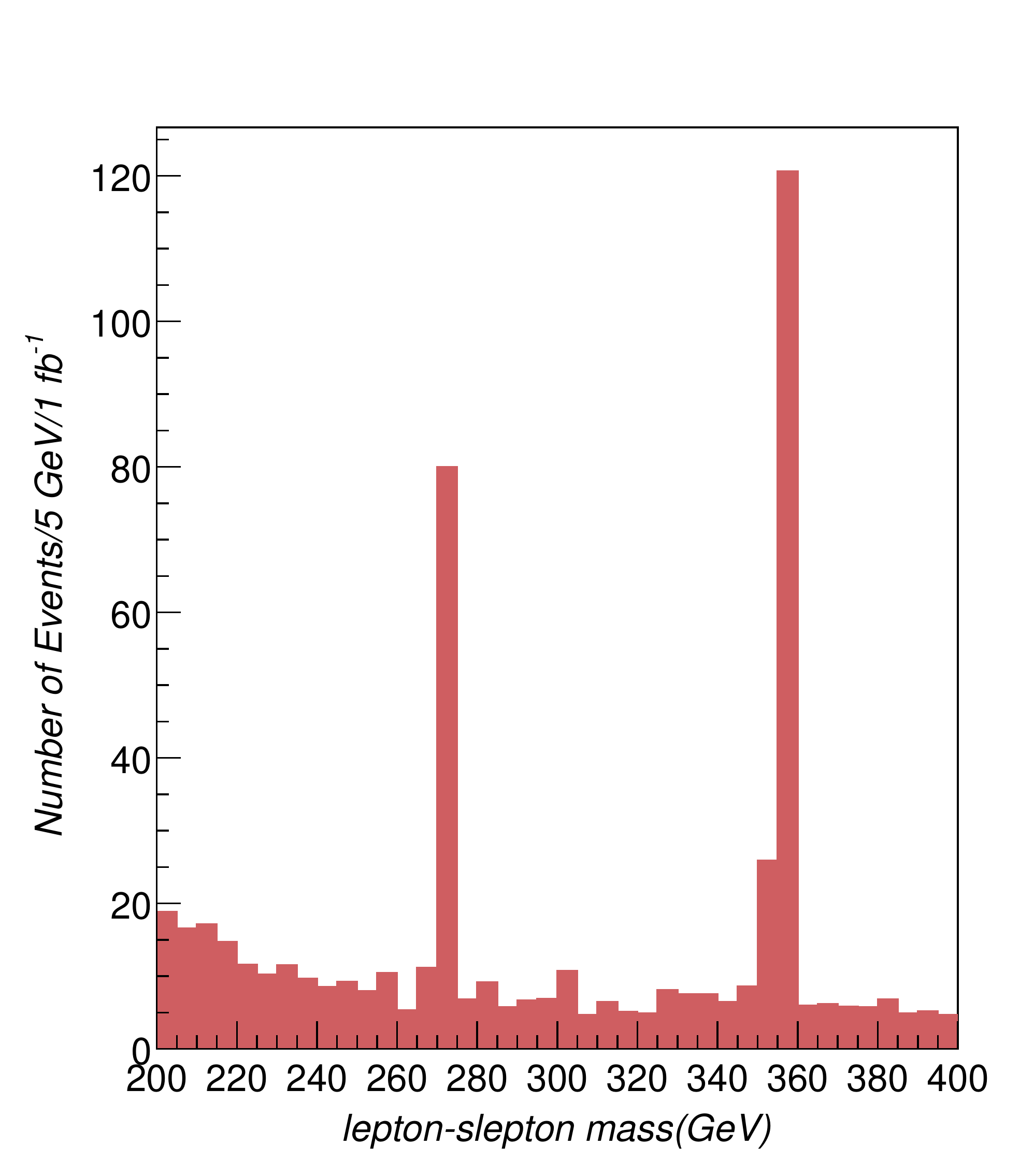}
 \includegraphics[width=4.8cm, height=5.5cm]{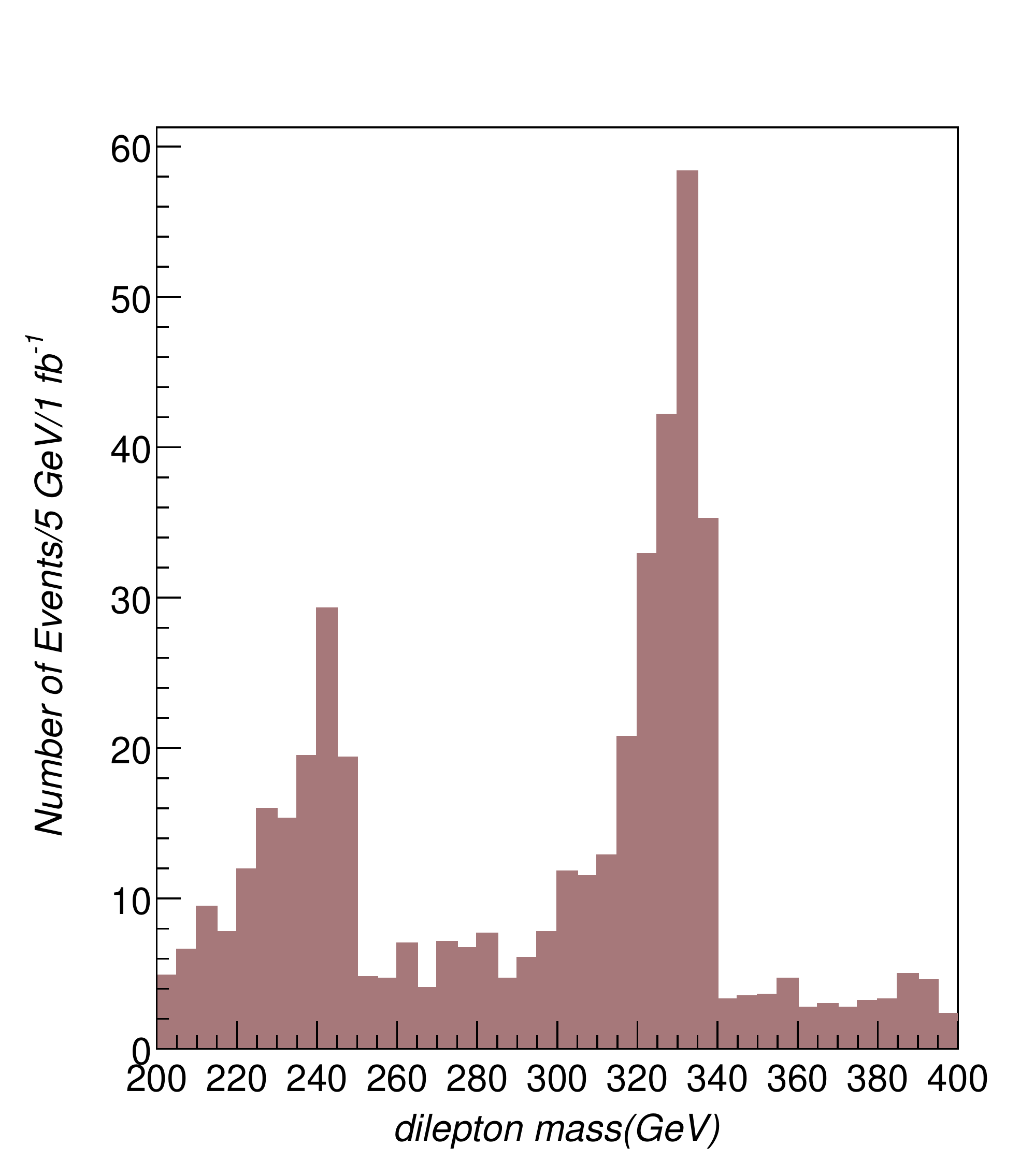}
  \includegraphics[width=4.8cm, height=5.5cm]{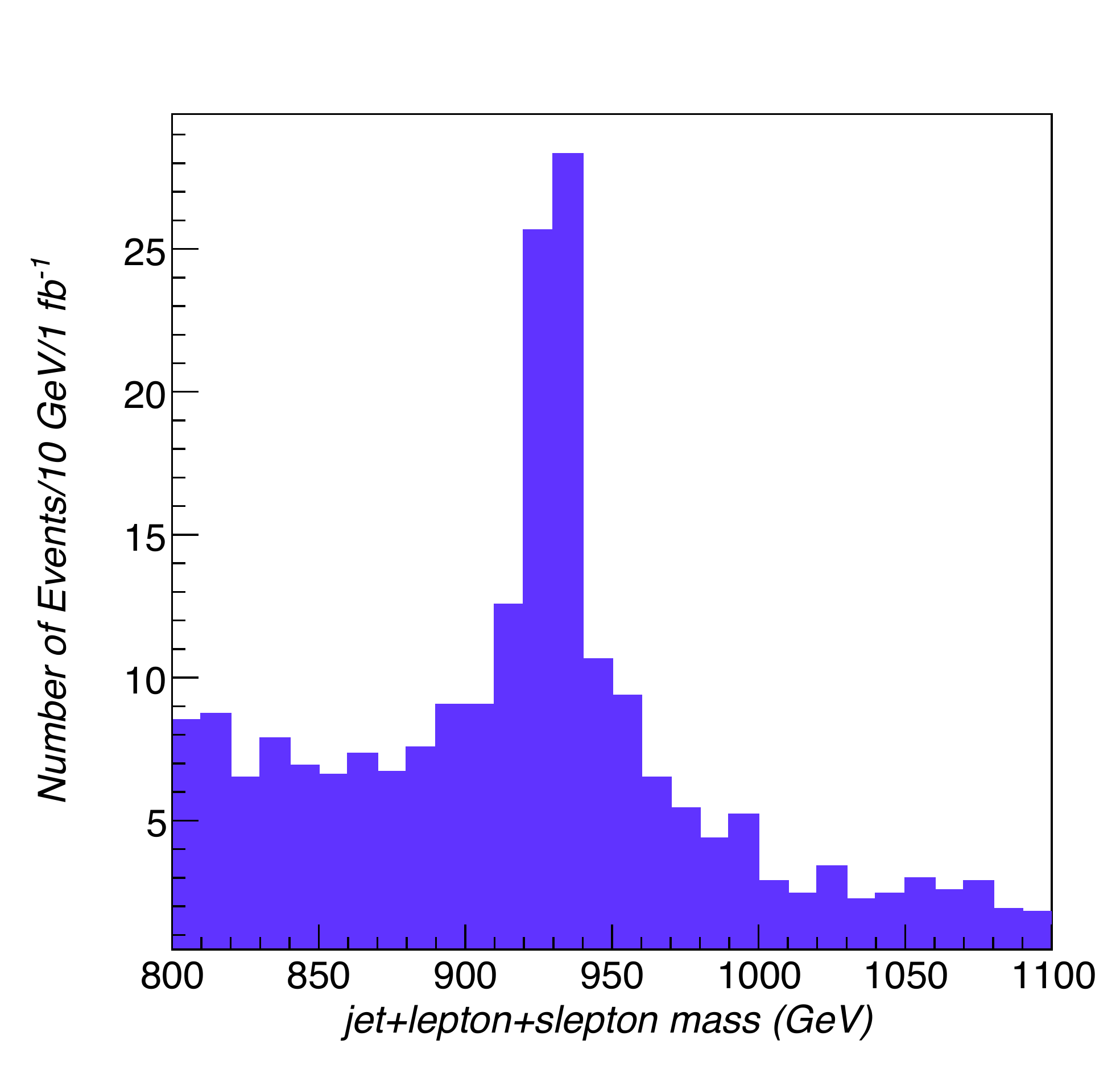}
 \caption{\textit{Left:}  Neutralino mass reconstruction from slepton-lepton invariant mass. \textit{Center:} Dilepton invariant mass distribution. \textit{Right:} Squark mass reconstruction. All plots are for $\sqrt{s}=14 $ TeV and ${\cal L}=1\,\rm{fb}^{-1}$.}
\label{2lepneu} 
\end{figure}

The slepton-lepton invariant mass distribution fully reconstructs the masses of the neutralinos $\tilde{\chi}^0_1$ and $\tilde{\chi}^0_3$. The left plot in Fig.~\ref{2lepneu} exhibits two clear peaks in the lepton-slepton invariant mass distribution, corresponding to $\tilde{\chi}_1^0$ and $\tilde{\chi}_3^0$.   On the other hand, if the  slepton are misidentified, they are  reconstructed as massless muons rather than true massive objects. 
and the dilepton invariant mass distribution shows edge-like structure.  
Both the shift of the maximum and the smearing of the distribution, apparent on the center plot in 
Fig.~\ref{2lepneu},
are the result of  this ``missing mass'' effect. 
The slepton+lepton+jet reconstruction determines $m_{\tilde{q}}$, see the right panel of Fig.~\ref{2lepneu}.

At 14 TeV and with 1 fb$^{-1}$ of integrated luminosity, we are able to reconstruct the masses of the neutralinos 
$\tilde\chi_1^0$ and $\tilde\chi_3^0$ with 96 and 132 clean events, respectively. For the squark we find 83 clean events.
In the case of 10 TeV and 200 pb$^{-1}$ these numbers get reduced by a factor of 13.
In the channels with five and six leptons one could further determine the masses of the heavy slepton, the sneutrino, the charginos and the other neutralinos (see Ref.~\cite{lepto} for details).

\subsection{Higgs discovery}

In leptogenic supersymmetry the transition $\tilde\ell_L\to h\tilde\ell_R$ is kinematically allowed and then the Higgs decays as $h\to b\bar{b}$.  Thus, the Higgs can get abundantly produced in cascade decays (as illustrated in the right panel of Fig.~\ref{decays}). The SM background is very efficiently reduced because of the high lepton multiplicity of the final state. Therefore, contrary to standard SUSY searches, it is possible to discover the Higgs through the analysis of a clean $b\bar{b}$ invariant mass distribution. 

The total cross-section for this process is 100 (320) fb at 10 (14) TeV, for our  sample spectrum.
We select events containing the Higgs asking for $n_{\ell}= 3,4, n_{\rm{jet}}\geq 4$ with cuts $|\eta_{\rm{\ell}, \rm{jet}}|<2.5, p_T^{\rm{\ell}}> 10$ GeV, $p_T^{\rm{jet}}>15$ GeV, and the isolation cuts $\Delta R_{\rm{jj}, \ell\ell, \rm{j}\ell}>0.4$. We order the jets in $p_T$ and
impose $p_T> 25$ GeV for the softest one. We then reconstruct the  invariant mass of the third and fourth jet and look for peaks. 
This very simple analysis has the advantage of not using $b$-tagging techniques, whose efficiency is not precisely known at the beginning.
We were able to estimate that, with 1 fb$^{-1}$ of data, 35 Higgs events  may be reconstructed. However, a more proper treatment of the combinatorial background and a refined analysis are needed to assess the actual LHC reach for this channel; this is under study by the ATLAS 
collaboration.


\section{Conclusions}
\label{concl}

Leptogenic supersymmetry is a well-motivated scenario where the ordering of the spectrum ensures, due to kinematics alone, copious lepton production in decay chains. The cascade decays of squarks pairs end up with collider-stable charged sleptons, so almost no missing energy is present, whereas the abundance of leptons renders the SM background  extremely small.

The discovery of a stable slepton is possible with the very first data. Most of the superpartner spectrum can be reconstructed 
 with 0.2 -- 0.4 fb$^{-1}$ at 10 TeV, for squark masses at the TeV.
The Higgs  discovery channel is  through the $h\to b \bar{b}$ decay in association with four leptons and two hard jets. We estimate the  prospects for Higgs discovery are good with less than 1 fb$^{-1}$ of data at 14 TeV, again assuming 1~TeV squarks.

The signals of leptogenic supersymmetry are very clean  and qualitatively different from standard SUSY searches, and are within the reach of the first run of LHC.


\begin{theacknowledgments}
This contribution is based on work done in collaboration 
with J.~Fan, V.~Sanz and W.~Skiba.
I acknowledge support from  the INFN ``Bruno Rossi'' Fellowship and from the U.S. 
Department of Energy (DoE) under contract No. DE-FG02-05ER41360.
\end{theacknowledgments}



\begin{thebibliography}{10}

\bibitem{lgm}
  A.~De Simone, J.~Fan, M.~Schmaltz and W.~Skiba,
  Phys.\ Rev.\  D {\bf 78}, 095010 (2008)
  [arXiv:0808.2052 [hep-ph]].


 \bibitem{gmsb}
For a review, see e.g.: G.~F.~Giudice and R.~Rattazzi,
 Phys.\ Rept.\  {\bf 322}, 419 (1999)
 [arXiv:hep-ph/9801271].

\bibitem{Fox:2002bu}
 P.~J.~Fox, A.~E.~Nelson and N.~Weiner,
 JHEP {\bf 0208}, 035 (2002)
 [arXiv:hep-ph/0206096];

\bibitem{lepto}
  A.~De Simone, J.~Fan, V.~Sanz and W.~Skiba,
  Phys.\ Rev.\  D {\bf 80} (2009) 035010
  [arXiv:0903.5305 [hep-ph]].
  
 
 \bibitem{cmstdr}
  G.~L.~Bayatian {\it et al.}  [CMS Collaboration],
 J.\ Phys.\ G {\bf 34}, 995 (2007).

 \bibitem{atlastdr}
G.~Aad {\it et al.}  [ATLAS Collaboration], ``ATLAS detector and physics performance. Technical design report.  Vol. 2,''
New: ``Expected Performance of the ATLAS Experiment : Detector, Trigger and Physics", arXiv:0901.0512 ; CERN-OPEN-2008-020.


\end{thebibliography}

\end{document}